version 20, dec. 2023

**Differential action of TIGIT on islet and peripheral nerve autoimmunity in the NOD mouse**

Chantal Bécourt, Sandrine Luce[1], Ute C. Rogner[*], Christian Boitard[*]

Institut Cochin

Inserm U1016 - CNRS UMR8104 - Université Paris Cité

123 boulevard Port Royal

bâtiment Cassini

F-75014 France

Telephone : +33176535582

Email: ute.rogner@inserm.fr ; christian.boitard@academie-medecine.fr

[*]equal corresponding and last authors

[1] Present address: Institut Mondor de Recherche Biomédicale (IMRB) - Inserm U955, Faculté de Médecine de Créteil, 8 rue du Général Sarrail, 94010 Créteil cedex

## Summary

We previously demonstrated that the abrogation of the ICOS pathway prevents type 1 diabetes development in the Non Obese Diabetic (NOD) mouse, but results in a $CD4^+$ T-cell dependent autoimmune neuromyopathy in aged mice. Pancreatic islet infiltrates in conventional NOD mice and neuromuscular infiltrates in *Icosl*$^{-/-}$ NOD mice have in common that they exhibit a strong enrichment in $CD4^+TIGIT^+$ T-cells, whilst TIGIT expression in the peripheral $CD4^+$ T-cells is limited to the $CD4^+FoxP3^+$ T-cell population.

When deleting *Tigit* on the NOD background, diabetes incidence was found increased. Peripheral $CD4^+CD226^+$ effector T-cells exhibited an increased frequency of IL-17 producing $CD4^+CD226^+RORgt^+$ T-cells *versus* a decreased frequency of IFNγ-producing $CD4^+CD226^+Tbet^+$ T-cells.

ICOS is expressed in both $CD4^+FoxP3^+$ and $CD4^+CD226^+$ splenic T-cell subsets. *Icosl* deletion leads to a decrease of $CD4^+FoxP3^+$ cells, with decrease of PD1 but increase of ICOS and CCRX3. Also in the *Icosl*$^{-/-}$ model, $CD4^+CD226^+$ T-cells are decreased by *Tigit* deletion, and showed an increase of $CD4^+CD226^+RORgt^+$ T-cells and a decrease of $CD4^+CD226^+Tbet^+$ T-cells.

However, deletion of *Tigit* in aged *Icosl*$^{-/-}$ NOD mice population did not increase the incidence of the autoimmune neuromyopathy observed in *Icosl*$^{-/-}$ NOD mice. Interestingly, the upregulation of $CD4^+CD226^+RORgt^+$ T-cells was partly rescued.

We conclude from our study that both *Icosl* and *Tigit* deletions on the NOD background lead to a shift between the ratio of IFNγ and IL-17-producing $CD4^+CD226^+$ effector cells. The ICOS-dependent neuromyopathy development remains dominant and is not further altered in the absence of TIGIT.

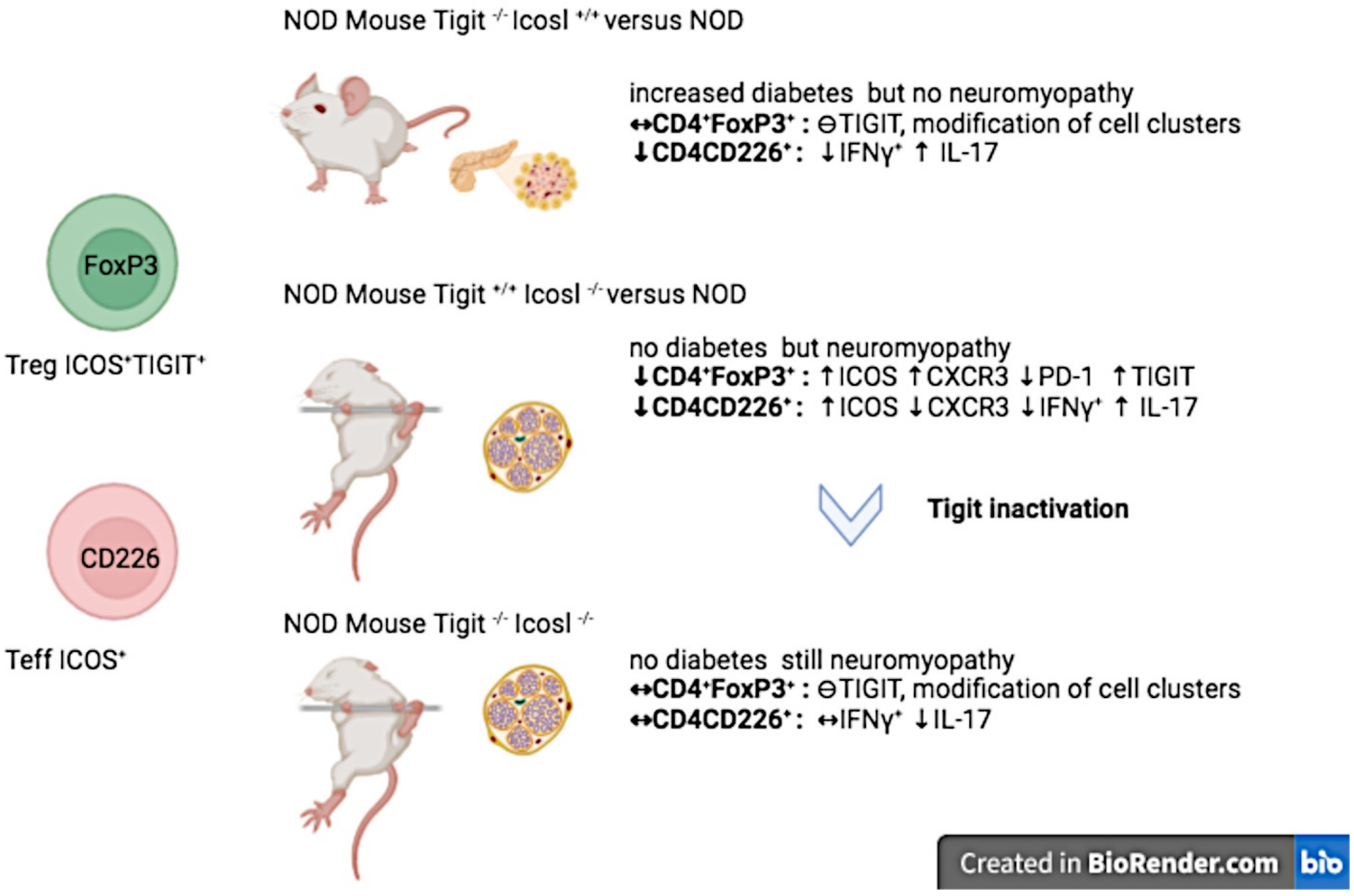

## Introduction

Immune checkpoints have emerged as key regulatory pathways in the activation and the inactivation of T lymphocytes by regulating T-cell receptor-driven signals. Beyond their role in maintaining T-cells unresponsive in cancer, they have been involved in controlling immune tolerance and its failure in major autoimmune diseases. Here, our focus is on the Non Obese Diabetic (NOD) mouse developing spontaneous type 1 diabetes (T1D) that is highly predominant in female mice. Several lines of evidence point to the role of regulatory T-cells (Tregs) and of immune checkpoint molecules in this model. The activation of $CD4^{+}$ T-cells against major islet cell autoantigens plays a major role in the development of the disease. It requires co-stimulatory signals in addition to TCR-mediated signals for the autoimmune response to proceed.

Immune checkpoints involve highly diversified families of co-signalling molecules at play in T-cell activation, expansion and differentiation. A fine balance between co-stimulatory and co-inhibitory receptors drives the regulation of immune responses. Their expression is highly diversified and their functions are tissue-dependent. Associations between autoimmune diseases and genes such as *CTLA4*, *PD1*, *TIM-3*, *TIGIT*, *LAG-3* (*HAVCR2*) gene polymorphisms have been reported (Anderson et al., 2016). In human, treatment of cancer with anti-CTLA4 and/or anti-PD1 monoclonal antibodies has shown the development in some patients of autoimmunity, including type 1 diabetes (Schnell et al., 2020). *In vivo* deletion of the *Cd28* and *B7* genes profoundly affects the development of diabetes in the NOD mouse (Hawiger et al., 2008; Jabs et al., 2002; Lenschow et al., 1996; Salomon et al., 2000). Seemingly, the deletion of either the *Cd28* related *Icos* (Hutloff et al., 1999) or the *Icosl* (Yoshinaga et al., 1999) genes prevents the development of type 1diabetes. In case of B7.2$^{-/-}$ (Cd82$^{-/-}$) (Louvet et al., 2009), *Icos*$^{-/-}$ and *Icosl*$^{-/-}$ NOD mice, an autoimmune shift from the islets towards peripheral nerves, and, in case of *Icos*$^{-/-}$/*Icosl*$^{-/-}$, mice in muscles, has been evidenced (Briet et al., 2017; Hawiger et al., 2008; Prevot et al., 2010).

By studying islet cell infiltrates in conventional NOD mice and neuromuscular infiltrates in *Icos*$^{-/-}$ NOD and *Icosl*$^{-/-}$ NOD mice, we observed a high frequency of T-cell ImmunoGlobulin and ITIM domain (TIGIT) expressing T-cells. TIGIT is expressed in activated T-cells as well as a subset of Treg cells. It interacts with two ligands, CD155 and CD112, as does the co-activator molecule CD226. Recently, it has been reported that CD226$^{-/-}$ NOD mice are protected from diabetes development (Shapiro et al., 2020). We aimed at studying the role of the TIGIT checkpoint molecule in the development of autoimmune disease. In order to evaluate the role of TIGIT in the development of autoimmunity, we deleted the *Tigit* gene using CRISPR-Cas9 in the conventional NOD background and introgressed the *Tigit* null mutation onto the *Icosl*$^{-/-}$ NOD line. *Tigit* deletion accelerated the development of diabetes in NOD mice, but it did not affect the kinetics of neuromuscular autoimmunity in *Tigit*$^{-/-}$*Icosl*$^{-/-}$ NOD mice. This indicates that TIGIT controls diabetes in the context of a functional ICOS pathway, but neither diabetes nor neuromuscular autoimmunity in the absence of the ICOS signalling pathway.

## Materials and methods

### Mice

All mice were bred and housed in our facilities under specific pathogen-free conditions. NOD WT and ICOSL$^{-/-}$ NOD mice were generated as described previously (Prevot et al., 2010). The prevalence of diabetes in our NOD WT colony reaches 10% in males and 60% in females by 6 months of age. All mice were maintained under specific and opportunist pathogen-free conditions and experimental studies were performed in accordance with the Institutional Animal Care and Use Guidelines and accredited by the Ethics Committee n°34 of Paris Descartes and the Ministry of Educational Superior and Research under the number APAFIS#16697-201711301425730v14.

### Invalidation of TIGIT on the NOD genetic background and introgression into the NOD ICOSL$^{-/-}$ line

CRISPR/Cas9 technology was used to invalidate the *Tigit* gene, in collaboration with TACGENE, Paris France for the design and selection of sgRNAs. The exon 2, coding for the transmembrane helix and cytoplasmic domain, was used as target sequence. Seven sgRNAs were designed (http://crispor.org) and synthesized by PCR for *in vitro* transcription with T7 RNA polymerase (Renaud et al., 2016). Two sgRNAs did not amplify and five were tested *in vitro* in mouse embryonic cells for the mutation rates using the T7 Endonuclease1 assay. The

sgRNA-550 (GAAGAACCGTGAAAAGGCACA**G)** and sgRNA-553 (GGTCTCACAACTGGACTATA**G**) were selected for a %indels of 10 and chosen for zygote injection and induction of the 1020 bp deletion within in murine *Tigit* gene.

sgRNAs were produced using the transcription T7 high yield kit (NEB) and purified with the EZNA micro-elute RNA clean up kit (Omega Bio-tek). Elutions were realized in MilliQ RNase free water and sgRNAs stored at -80°C. Overlapping PCR fragments were then analysed by sequencing. The CAS9-3NLS protein was produced by TACGENE. Transgenesis was performed by the PHRTEC platform at the Cochin Institute. sgRNA/Cas9 complex formation was realized by denaturing sgRNA at 80°C for 2 min, stored on ice for 2 min before adding the protein using a 1-1.5:1 molar stoichiometry ratio (sgRNA/Cas9-protein). Complexes were incubated 10 min at room temperature and kept on ice prior to dilution in a buffer containing 20 mM Hepes-NaOH pH 7.5, 150 mM-250 mM KCl. We performed one cell embryo injection using the diluted sgRNA/Cas9 protein complexes in fertilized eggs obtained from super-ovulated female wildtype NOD mice, previously mated with male NOD mice.

Injected 2-cell stage eggs were incubated O/N at 37°C and re-implanted into female SWISS mice. Genomic DNA was extracted from tail-biopsies of transgenic mice using proteinase K and SDS buffer at 56°C, O/N. DNA was precipitated using saturated NaCl and isopropanol, and resuspended in TE-4 buffer at 100 ng/µl. Genotyping PCR reactions were realized using the following primers: Tigit-KO-F (5'-tatgaccaggaaggggtcag-3') and Tigit-KO-R (5'-gtgcagccaatgagttctga-3') with an annealing temperature at 60 °C.

The expected size for wildtype (WT) alleles was 1651bp and for the knockout (KO) allele 597 bp (Figures 3,4). For identified KO animals, we subcloned the PCR product in the pCR4-TOPO vector (Invitrogen) for sequencing analysis to confirm the complete invalidation of *Tigit* gene in NOD TIGIT$^{-/-}$ animals using SnapGene software (UNC Center for Bioinformatics). Flow cytometry analysis was performed on the spleen and pancreatic infiltrating cells of *Tigit*$^{-/-}$ animals to confirm the absence of TIGIT expression on TCR$^{+}$CD4$^{+}$FoxP3$^{+}$ cells.

**Genetic analysis**

Genomic DNA extracted from mouse tail tip biopsis using standard protocols was processed and hybridized on Affymetrix Mouse Diversity Genotyping Arrays (Santa Clara, California, USA) according to the manufacturer's instructions. For data extraction, genome coordinates were determined using the assemblies UCSC version mm10 and NCBI version GRCm38. Non-informative markers were manually removed for further comparative strain analysis.

The primer sequences to detect the *Icosl* mutation (product size 300bp *versus* 225bp) were as follows:

B7-1 5'-GTT TGC TAC AAG ACA CTA ACG AG-3'
B7-2 5'-CAG ACC ACT CAA GTT GAA ATG G-3'
B7-3 5'-GGC AGC CTG TTT GGA AGA AGC TCC-3'
B7-4 5'-CGA ATT CGC CAA TGA CAA GAC GCT GG-3'

**Diabetes and neuromyopathy assessments**

Diabetes was assessed by monitoring mice as described previously (Prevot et al., 2010). Neuromyopathy was identified by observing the mice weakly for early signs of disease development, such as extension of the legs and shivering.

**Histological analysis**

Pancreata were excised, fixed in 70% ethanol and processed for paraffin embedding. Three sections (5 µm) taken at 100 µm intervals, stained using hematein-eosin, and visually analysed.

**Flow cytometry analysis**

**Antibodies**

The following antibodies (Abs) were used, hybridoma clones are indicated in parentheses: anti-CD45 Alexa Fluor 700 (30-F11), anti-TCR APC-eFluor 780 (H57-597), anti-CD8 PE-Cy7 (53-6-7), anti-CD8 AF700 (53-6-7), anti-CD4 PE (GK1-5), anti-TIGIT PE (GIG07), anti-PD1 PerCP-eFluor 710 (J43), anti-ICOS Biotin (7E 17G9), anti-CXCR3 APC (CXCR3-173), anti-FOXP3 Percp-Cyanine 5-5 (FJK-16s), anti-FOXP3 Alexa Fluor 488 (FJK-16s), anti-TNF Alpha eFluor 450 (MP6-XT22), anti-IFNγ Alexa Fluor 700 (XMG1-2), anti-T-Bet PE (4810), anti-ROR gamma(t)PE-Efluor 610 (B2D) all from eBioscience. anti-CD45R/B220 V500 (RAS-6B2), anti-CD4 BV786 (GK1-5), anti-IL-17 BV788 (TC11-18H10), Streptavidine BV650 from BD Bioscience, anti-CD226 BV421 (TX42-1), anti-CD226 BV605 (TX42-1) from Biolegend.

**Immunofluorescence staining**

was performed as described previously (Prevot et al., 2010). Intracellular cytokine staining was performed according to manufacturer's instructions for staining intracellular antigens (eBioscience) or using the True-Nuclear Transcription Factor Buffer Set (BioLegend). Flow cytometric analysis was performed using a FACS FORTESSA and FlowJo softwares (Tree Star, Ashland, OR). Living cells were gated on side scatter versus forward scatter density plots.

**Cytokine production analysis**

$2x10^6$ to $4x10^6$ cells/ml from $ICOSL^{+/+}$, $ICOSL^{-/-}$, $TIGIT^{-/-}$ and $TIGIT^{-/-}ICOSL^{-/-}$ NOD mice were incubated 4 hours at 37°C with Phorbol 12-myristate 13-acetate (PMA 10 ng/ml; Sigma), ionomycin (500 ng/ml; Sigma) and brefeldine A (10 µg/ml; Sigma). Cells were then collected, washed and tested for IFNγ, TNF alpha and IL-17 production by flow cytometry, gated on T cell populations.

**Isolation and purification of immune cells**

a) Muscle infiltrate: The muscles were removed from the limbs of the mice and placed into a 10 cm tissue culture dish containing 10 ml of HBSS with 1% glutamine. Then the bones and tendons were removed and the preparation was minced thoroughly, transferred in 50 ml conical tube and centrifuged 2000 rpm for 3 min. The supernatant was discarded and an equivalent volume of 0.2% collagenase type II (Worthington Biochemicals) was added and incubated in a 37°C water bath for 60 min, with mixing well every 10 min After the sample was triturated with 5 aliquots of complete DMEM and centrifuged, the suspension was passed through a 100 µm filter mash. After centrifugation, the cell suspension was gently overlaid onto a Percoll (Amersham Pharmacia Biotech AB) gradient and centrifuged at 2500 rpm for 20 min without break at 25 °C. The cells were removed from the 70%, 40% Percoll interface and transferred to a fresh conical tube to wash out the Percoll. These cells were maintained overnight à 37°C, 5% CO2 in RPMI with 10% FCS.

b) Nerve infiltrate: Sciatic and brachial nerves were cut into small fragments and incubated in RPMI 1640 supplemented with 1.6 mg/ml collagenase (type IV Worthington, LS004188) and 200 µg/ml DNase I (Roche) at 37°C for 30 min. Cells were dissociated by repeated pipetting, incubated at 37°C for 10 min, and washed. The percoll gradient, washing and incubation was performed as described under a.

c) Pancreas islets infiltrate: After infusion of the pancreas via the common bile duct with 5 ml of collagenase 1 mg/ml (type XI, Sigma C7657), they were removed and placed in RPMI 1640 supplemented with 2 ml of collagenase solution and digested for 7 min using a 37°C water bath. Then after having carried out a pre-isolation on a 40%-23%-20%-11% ficoll gradient (Sigma Ficoll type 400, F4375), the islets were hand picked and maintained overnight à 37°C, 5% $CO_2$ in RPMI with 10% FCS.

**2.7 Statistical analysis**

Appropriate statistical tests (log rank, Mann-Whitney) were performed as indicated in the figure legends using GraphPad Prism Version 4.0b software (GraphPad Software, La Jola, CA).

**2.8 Ethical approval**

This study was carried out in accordance to the recommendations of Institutional Animal Care and Use Guidelines. The protocol was approved by the ethic committee number CEEA34.CB.024.11.

## RESULTS

### ICOS marks peripheral regulatory and effector CD4$^+$ T-cell populations in NOD mice

NOD mice spontaneously develop insulin-dependent diabetes after 10-12 weeks of age. The ICOS pathway contributes to the diabetes process (Hawiger et al., 2008). We showed that *Icos*$^{-/-}$ and *Icosl*$^{-/-}$ NOD mice were protected from diabetes, but developed an auto-immune neuro-myopathy with predominance in female mice after 26 weeks of age (Briet et al., 2017; Prevot et al., 2010). Our data further showed that, whereas acceleration of diabetes in BDC2.5 ICOS$^{-/-}$ NOD mice related with a dominant defect in regulatory T-cells, protection from diabetes results from defective activation of autoimmune diabetogenic effector T-cells in ICOS$^{-/-}$ NOD mice. *Icos*$^{-/-}$ and *Icosl*$^{-/-}$ NOD mice exhibit the same phenotype. The non-NOD interval around the *Icos* gene introgressed onto the NOD genetic background in *Icos*$^{-/-}$ NOD mice overlaps with diabetes protecting *Idd* 5.1, 5.2 and 5.3 alleles. In *Icosl*$^{-/-}$ NOD mice a much smaller interval around the *Icosl* gene, devoid of *Idd* genes, was introgressed (Briet et al., 2017). Thus, *Icosl*$^{-/-}$ NOD mice were selected for further studying autoimmune neuromyopathy in this report. This model also has the advantage that it allows to follow ICOS as a marker of immune cells.

To better define the expression domains of ICOS and other co-activation markers in CD4$^+$ T-cells, we first analysed lymphocytes in the peripheral immune system (spleen) of female NOD WT mice (n=11) at 12 weeks of age (Supplementay Figure S1). ICOS was almost not expressed by CD4$^+$FoxP3$^-$ and CD8$^+$ T-cells (Figure S1c,d). CD4$^+$FoxP3$^+$ T-cells represented 11.9% to 15.9%, and CD4$^+$FoxP3$^-$CD226$^+$ cells 2.5% to 5.3% of the CD4$^+$ T-cells. Among CD4$^+$FoxP3$^+$ and CD4$^+$FoxP3$^-$CD226$^+$ T-cells, 22.4% to 36% and 41.2% to 55.6%, respectively, expressed ICOS. ICOS was mainly expressed by CD4$^+$FoxP3$^+$ T-cells that expressed the co-inhibitory molecule TIGIT (20.9% to 27.9%), the immune regulator PD1 (9.4% to 14.8%) and the homing marker CXCR3 (13.6% to 20.7%), and by CD4$^+$FoxP3$^-$CD226$^+$ effector T-cells that expressed CXCR3 (29.3% to 58.3%) (Figure S1b,c). As a comparison, ICOS was only detected on 1% to 2.6% (Figure 1S) and TIGIT only on 0.4-1.2% of CD4$^+$FoxP3$^-$ T cells (n=6).

The expression pattern was confirmed in older mice at 16 weeks (Figure S2) and 24 weeks of age (Figure 1).

## The ICOS pathway controls the peripheral $CD4^+$ Th17/Th1 balance in the NOD mouse

We analysed the expression of ICOS and other T-cell co-activation markers in lymphocytes obtained from the spleens of wild type and $ICOSL^{-/-}$ 24-30 week-old female NOD mice (Figure 1). A mild decrease in $TCR^+$ cells and $CD8^+$ T-cells was observed in $ICOSL^{-/-}$ mice (Figure 1a). As previously reported (Prevot et al., 2010), a major decrease in $CD4^+FOXP3^+$ T-cells was observed in $ICOSL^{-/-}$ mice as compared to wild type NOD mice (Figure 1a). Meanwhile, $ICOSL^{-/-}$ NOD mice showed a significant increase in $CD4^+FOXP3^+ICOS^+$ T-cells, and, even more strikingly, of $CD4^+FOXP3^+$ T-cells that expressed TIGIT, CXCR3 and co-expressed ICOS and TIGIT or ICOS and CXCR3 (Figure 1b,d,e). The decrease in $CD4^+FoxP3^+$ T-cells observed in *Icosl*$^{-/-}$ as compared to wild type NOD mice was paralleled by a decrease in $CD4^+FoxP3^+$ T-cells that co-expressed PD1. $CD4^+FOXP3^+PD1^+$ T-cells were decreased by 58%, and $CD4^+FOXP3^+PD1^+TIGIT^+$ T-cells by 54% in $ICOSL^{-/-}$ as compared to wild type NOD mice. A decrease was also observed in $CD4^+FOXP3^-CD226^+$ T-cells in $ICOSL^{-/-}$ as compared to wild type NOD mice, along with a decrease in $CD4^+CD226^+CXCR3^+$ T-cells (Figure 1c). A decrease in IFNγ- and TNF- and an increase in IL-17-expressing $CD4^+CD226^+$ T-cells was observed in $ICOSL^{-/-}$ as compared to wild type NOD mice (Figure 1c,g), as well as an increase in $CD4^+CD226^+$ T-cells that expressed TBET or RORγT (Figure 1c,f,g). When analyzing *Icosl*$^{-/-}$ NOD mice before neuromyopathy onset at 16 weeks, the decrease in $CD4^+Foxp3^+$ T-cells, $CD4^+FoxP3^+PD^{1+}$ and $CD4^+FoxP3^-CD226^+$ T-cells was already observed, as well the increase of IL-17$^+$ $CD4^+FoxP3^-CD226^+RORγt^+$ and $CD4^+FoxP3^-CD226^+IL\text{-}17^+$ cells (Figure S2).

We therefore concluded from these results that the *Icosl* mutation led to a decrease in Treg and Teff cell percentages amongst $CD4^+$ T-cells, along with an increase of IL-17 and decrease in IFNγ- and TNF- producing cells.

## $ICOS^+$ and $TIGIT^+$ $CD4^+$ T-cells are enriched in pancreatic, muscular and nerve infiltrates

We next analysed pancreatic infiltrates of 24 week old NOD WT mice and nerve and muscle infiltrates of 28-32 and 25-28 week old, respectively, NOD *Icosl*$^{-/-}$ mice as summarized in Figure 2. Additional data for the pancreatic infiltrate of female wildtype mice at 12 weeks of age are summarized in Figure S1. The infiltrates contained various proportions of $TCR^+CD4^+$ T-cells, including 11.3 to 18 % $CD4^+FopxP3^+$ T-cells (Figure 2e). Among $CD4^+FoxP3^+$ T-cells, a majority of infiltrating cells expressed ICOS (>80 %), but also TIGIT (>65%), depending on the tissue obtained from NOD WT mice or NOD *Icosl*$^{-/-}$ mice (Figure 2a,b,c). Interestingly, compared to the periphery, CD4$^{+}$FoxP3$^{-}$ T-cells also expressed ICOS (> 54%) and TIGIT (>29%) (Figure 2c,g). We noticed in all infiltrates a strong increase of CD4$^{+}$ICOS$^{+}$TIGIT$^{+}$ T-cells as compared to the spleen, that concerned both CD4$^{+}$FoxP3$^{+}$ (>64%) and CD4$^{+}$FoxP3$^{-}$ (>28%) T-cell populations (Figure 2e). Interestingly, in infiltrates, CD8$^{+}$ T-cells also express ICOS (>17%) and TIGIT (>32%).

*CD4$^{+}$FoxP3$^{+}$ T-cell frequencies in tissue infiltrates compared to spleen*

In the spleen of NOD WT mice, frequencies of regulatory CD4$^{+}$FoxP3$^{+}$ T-cells comprised 11.9% to 15.9% of the CD4$^{+}$ T-cell population at 12 weeks of age (WT females, n=11), and 9.7% to 20.6% at 24-30 weeks of age (n=22) (Figure 1 and S1). In pancreatic infiltrates recovered from 24 week-old WT females CD4$^{+}$FoxP3$^{+}$ cells constituted 16.5% to 21.8% of CD4$^{+}$ T-cells (9.4% to 16% at 12 weeks), in infiltrates from *Icosl*$^{-/-}$ females (> 25 weeks) in nerves 9.4% to 14.2%, and in muscles 14.6% to 17.7% (Figure 2).

*CD4$^{+}$ICOS$^{+}$ T-cell frequencies in tissue infiltrates compared to spleen*

In WT NOD mice, 22.4% to 36% CD4$^{+}$FoxP3$^{+}$ peripheral regulatory T-cells expressed ICOS in 12 week-old female WT NOD mice (n=11), a percentage that varied from 17.3% to 45.1% at 24-30 weeks of age (n=22). CD4$^{+}$FoxP3$^{-}$CD226$^{+}$ T-cells, that constituted about 2.5% to 5.3% of CD4$^{+}$ cells at 12 weeks of age (n=11) and 3.8% to 19.5% at 24-30 weeks of age (n=22), also expressed ICOS in the periphery (41.2% to 55.6%, n=11, at 12 weeks of age; 50.6% to 75.9% at 24-30 weeks, n=22) (Figures 1 and S1).

In infiltrates, ICOS was expressed on both CD4$^{+}$FoxP3$^{+}$ and CD4$^{+}$FoxP3$^{-}$ T-cells. In pancreatic infiltrates at 12 weeks of age, ICOS was expressed on 59.9% to 81.5% of CD4$^{+}$Foxp3$^{+}$ T-cells, and on 39.4% to 60.6% of CD4$^{+}$Foxp3$^{-}$ T-cells (Figure S1). At 24 weeks we found 72.4% to 93.1% of CD4$^{+}$FoxP3$^{+}$ T- cells and 47% to 61.2% of CD4$^{+}$FoxP3$^{-}$ T-cells ICOS positive (Figure 2). Interestingly, in *Icosl*$^{-/-}$ mice, percentages varied in nerve infiltrates from 89.4% to 97.7% of CD4$^{+}$Foxp3$^{+}$ and 83.8% to 96.2% of the CD4$^{+}$Foxp3$^{-}$ T-cells. In muscle infiltrates, we found 95 to 97.1% and 59.4 to 64.5%, respectively.

Within the CD4$^{+}$Foxp3$^{-}$ T-cell population it should be noted that CD4$^{+}$FoxP3$^{-}$CD226$^{+}$ cells were almost absent from the infiltrates (<1%), although CD4$^{+}$FoxP3$^{-}$CD226$^{+}$ effector cells represented 2.5% to 5.3% (WT females, 12 weeks, n=11) and 3.8% to 19.5% (WT females, 24-30 weeks, n=22) in spleen.

*$CD4^+$ $TIGIT^+$ T-cells are enriched in tissue infiltrates compared to spleen*

When further analysing these cell populations in our different mouse models, we realized that, while $CD4^+TIGIT^+$ are only present at relative low percentages in the periphery of NOD WT mice, they are highly enriched in the tissue infiltrates during disease development. In spleens, $TIGIT^+$ cells comprise 20.9% to 27.9% $CD4^+FoxP3^+$ T-cells (n=11) in 12 week-old female NOD mice, and 23.6% to 47.9% (n=22) in 24 week-old non-diabetic female NOD mice (Figure 2 and S1). As a comparison, TIGIT is detected on 0.4-1.2% $CD4^+FoxP3^-$ T-cells at 12 weeks (n=6) and 0.7% to 1.6% $CD4^+FoxP3^-$ T-cells at 24 weeks (n=10). Similar findings were made in *Icosl*$^{-/-}$ female mice at 28 weeks of age (29.1% to 48.7% in the $CD4^+FoxP3^+$ T-cells, and 1.3% to 4% in the $CD4^+FoxP3^-$ T-cells, n=11) (Supplementary Table 1).

In pancreatic infiltrates of wild type 12 week-old female NOD mice, the percentages of $TIGIT^+$ T-cells varied from 38.8% to 68.5% in the $CD4^+FoxP3^+$ population and from 23.6% to 39.3% in the $CD4^+FoxP3^-$ population (Figure S1). At 24 weeks of age we found 53.3% to 77.5% in the $CD4^+FoxP3^+$ and 23.9% to 32.7% in the $CD4^+FoxP3^-$ population, respectively (Figure 2). In nerve infiltrates of the around 30 weeks old *Icosl*$^{-/-}$ female mice, 65.9% to 73.6% of the $CD4^+FoxP3^+$ T-cell population, and 33.1% to 39.7% of the $CD4^+FoxP3^-$ T-cell population (Figure 2) expressed TIGIT. Similar findings were made in muscle infiltrates, $TIGIT^+$ cells comprised about 77.1% to 77.7% of the $CD4^+FoxP3^+$ T-cells, and 26.8% to 41% of the $CD4^+FoxP3^-$ T-cells (Figure 2).

The data suggest a strong increase of $TIGIT^+$ T-cells within infiltrates that is particularly observed in the $CD4^+FoxP3^-$ T-cell subset. The expression of TIGIT on regulatory and effector T-cells and their increase within infiltrates suggested that it plays an important role in disease development. We therefore decided to analyse the effect of the *Tigit* mutation in both WT NOD mice and *Icosl*-deficient NOD mice.

**Increased incidence of diabetes but not neuromyopathy in TIGIT$^{-/-}$ NOD mouse models**

To answer the question of whether TIGIT can modulate diabetes development, we constructed a knockout mouse model using the CRISPR/Cas9 technology. The experiment was designed to remove the exon 2 of *Tigit*, resulting in a truncated protein sequence that ends at amino acid 130, and does no longer contain the transmembrane helix and the cytoplasmic domain of TIGIT. Three transgenesis experiments using NOD mice resulted in a total of 39 offsprings. The *Tigit* deletion was identified in three mice (7.7%) by PCR genotyping. By sequence analysis we observed a homozygous deletion of the exon 2 (Figure 3a) for mice 1 and 2 affecting the protein expression in splenocytes (Figure 3a), whilst animal 36 carried a heterozygote mutation. The *Tigit*$^{-/-}$ mice were then mated to NOD WT mice to establish the new mouse strain. To confirm the absence of TIGIT, we also studied the expression on regulatory T-cells from pancreatic infiltrates by flow cytometry. We observed a complete absence of TIGIT expression on TCR$^{+}$CD4$^{+}$Foxp3$^{+}$ pancreatic infiltrating T-cells in the offsprings of TIGIT$^{-/-}$ NOD mice *versus* NOD WT mice (63% of TIGIT expression in TCR$^{+}$CD4$^{+}$Foxp3$^{+}$ T-cells).

As shown in figure 3, TIGIT$^{-/-}$ NOD mice showed an increased incidence of diabetes in both male and female mice. In control TIGIT$^{+/+}$ NOD mice, 46% of females and 10% of male became diabetic by 40 weeks of age. By contrast, 85% of TIGIT$^{-/-}$ female and 41% of TIGIT$^{-/-}$ male mice became diabetic by 40 weeks of age. Diabetes onset occurred earlier in male TIGIT$^{-/-}$ NOD mice than in male TIGIT$^{+/+}$ NOD mice. The first TIGIT$^{-/-}$ NOD mice that developed diabetes were 13 weeks old, as compared with 26 weeks of age in the case of TIGIT$^{+/+}$ NOD mice. In TIGIT$^{-/-}$ NOD mice, diabetes onset occurred earlier in males than in females, onset spanned from 17 to 31 weeks of age in female mice and from 13 to 29 weeks in male mice. Altogether, these results suggest that the lack of TIGIT expression leads to an increased incidence of diabetes. Whilst pancreata from diabetic mice were highly infiltrated, pancreata obtained from pre-diabetic 12-week-old TIGIT$^{-/-}$ NOD mice showed only slightly more islet infiltration compared to pancreata from TIGIT$^{+/+}$ NOD mice. Six female *Tigit*$^{-/-}$ NOD mice showed insulitis in 52% of islets (58 islets), compared to six female *Tigit*$^{+/+}$ NOD mice showing 46% infiltrated islets (65 islets). TIGIT$^{-/-}$ NOD mice male mice at 12 weeks of age showed 38% infiltrated islets (82 islets) compared to 25% (68 islets) in TIGIT$^{+/+}$ NOD mice males ($P$ < 0.1 in Chi-square test). As an increased incidence of diabetes was observed in TIGIT$^{-/-}$ NOD mice, a key issue was whether the same increase in autoimmune development was observed in *Icosl*$^{-/-}$ NOD mice. As shown in figure 4, no significant difference was observed in the development of neuromyopathy in *Icosl*$^{-/-}$ *Tigit*$^{-/-}$ NOD mice as compared to *Icosl*$^{-/-}$ NOD mice or *Icosl*$^{+/+}$ NOD mice. Meanwhile, *Icosl*$^{-/-}$ *Tigit*$^{-/-}$ NOD mice remained diabetes-free.

We therefore conclude that the TIGIT pathway can modulate diabetes development but not that of neuromyopathy in the absence of a functional ICOS pathway.

We next investigated if the T-cell population in tissue infiltrating cells was altered upon introduction of the *Tigit* mutation in the NOD WT mice. (Supplementary Table 1).

When comparing female WT mice to *Tigit*$^{-/-}$ mice no significant changes were observed. In particular, the null mutation of *Tigit* did not influence the number of ICOS$^{+}$ T-cells (2-4 comparative experiments at 12 weeks (Supplementary Table 1). The analysis of nerve infiltrates obtained from *Icosl*$^{-/-}$*Tigit*$^{-/-}$ NOD female mice at 28-33 weeks of age (Figure 2) showed similar T-cell numbers as in *Icosl*$^{-/-}$*Tigit*$^{+/+}$ as compared to *Icosl*$^{-/-}$*Tigit*$^{-/-}$ NOD mice (63.5% CD4$^{+}$, 31.6$^{+}$ CD8$^{+}$, 11.3%, CD4$^{+}$FoxP3$^{+}$ versus 51.3% CD4$^{+}$, 40% CD8$^{+}$, 10.4% CD4$^{+}$FoxP3$^{+}$, respectively). The number of ICOS$^{+}$ T-cells was not significantly altered by the null mutation of *Tigit* (94.47% of CD4$^{+}$Fopx3$^{+}$ICOS$^{+}$ T-cells versus 97.25%, and 89.73% of CD4$^{+}$Fopx3$^{-}$ICOS$^{+}$ T-cells versus 97.15%, respectively), in the ICOSL$^{-/-}$ strain versus the ICOSL$^{-/-}$ TIGIT$^{-/-}$ strain.

We concluded that the T-cell composition of the infiltrates was not strongly altered in the absence of TIGIT. We therefore asked the question if the peripheral cellular compartment represents alterations that could be involved in increased diabetes incidence.

**TIGIT controls the peripheral Th1/Th17  NOD mice balance in the NOD mouse**

We first compared female *Tigit*$^{-/-}$ NOD versus wild type NOD mice before diabetes onset at 12 weeks of age (Figure 3). We found no significant changes in T-cells or B-cells and only a small decreased percentage of CD4$^{+}$FoxP3$^{+}$ cells from an average of 12.4% ±1.2% to 11.5% ±1.2% (WT n=12, KO n=13). The CD4$^{+}$CD226$^{+}$ T-cell population remained stable. Interestingly, the number of CD4$^{+}$CD226$^{+}$ Tbet and CD4$^{+}$CD226$^{+}$ RORgt expressing cells changed. The decrease of Tbet expressing cells (on average 21.2% ±8.7 versus 11.4% ±6.6%) was associated with a decrease of IFNγ-expressing cells (33.2% ±6.5% versus 28.9% ±2.8%), and the increase of RORγt expressing cells (2.4% ±0.8% versus 3.2% ±0.9) with an increase in IL-17 (1.3 ±0.7% versus 1.5% ±0.5) producing cells.

We attempted to validate this finding in an experiment at 16 weeks of age using five WT and six *Tigit*$^{-/-}$ mice (Figure S3). Interestingly, a change in CD4$^{+}$FoxP3$^{+}$ T-cells was again not observed. Most interestingly, RORgt expressing cells were then not increased amongst the CD4$^{+}$FoxP3$^{-}$CD226$^{+}$ cells (P=0.23404), and IL-17 producing cells were even decreased (P=0.01046). This indicates that the effect of *Tigit* mutation could be age-dependent.

As shown in figure 4, at 28 weeks of age no difference was observed in TCR$^{+}$, CD4$^{+}$, CD8$^{+}$, CD220$^{+}$, CD4$^{+}$FoxP3$^{+}$, CD4$^{+}$CD226$^{+}$ cell distributions between *Icosl*$^{-/-}$ (n=13) and *Icosl*$^{-/-}$*Tigit*$^{-/-}$ (n=17) NOD mice. A strong downregulation of CD4$^{+}$FoxP3$^{-}$CD226$^{+}$ RORγt expressing cells was found for the ICOSL$^{-/-}$TIGIT$^{-/-}$ mice, confirmed by the analysis of IL-17 expression. This suggests that the effect of ICOSL mutation on the IL-17 producing CD4$^{+}$FoxP3$^{-}$CD226$^{+}$ T-cells was reversed by the *Tigit* mutation (Figure 4g).

## DISCUSSION

*Loss of Tigit affects type 1 diabetes but not neuromyopathy*

In this study we found that ICOS and TIGIT are expressed on large proportions of T cells in infiltrated tissues (pancreas, nerves, muscles) during autoimmunity in the NOD mouse. Interestingly, Spence et al. (Spence et al., 2018) previously described that Tregs in the islets showed evidence of antigen stimulation, correlating with higher proliferation and expression of the activation markers ICOS, and TIGIT. They demonstrated that the specificities of Tregs in a natural repertoire are crucial for opposing the progression of autoimmune diabetes. These discoveries raised the question of *Tigit* deletion would impact on autoimmune disease development. Indeed, *Tigit* deletion clearly increased diabetes development in our WT nonobese diabetic mice. The *Icos* pathway remained however dominant over the *Tigit* pathway in the context of the $CD4^+$ T cell dependent neuromyopathy. *Tigit* deletion strongly alters the Treg cell clusters, and both molecules clearly regulated $CD4^+CD226^+$ effector cell frequency and function in the peripheral immune system. Striking was the loss of Tregs and amongst them the loss of PD1 expressing cells in *Icosl* deficient mice. This regulation may well affect the outcome of the *Tigit* mutation in the *Icos* pathway deficient background.

*ICOS pathway in type 1 diabetes: peripheral regulation of PD1, TIGIT and CXCR3 in Tregs*

Previously it has been described that Programmed cell death 1 (PD1) deficient Tregs have increased suppressive activity (Zhang et al., 2016), which would correlate with our finding that a defective ICOS pathway still protects from diabetes despite the loss of Tregs in the periphery.

In addition the loss of PD1 we found increase of TIGIT expressing Tregs in the periphery of *Icosl*$^{-/-}$ mice. We also observed strong modifications of the $CD4^+FoxP3^+$ cell clusters in absence of TIGIT (Supplementary figure 4). Such finding suggests more functional modifications of the Treg populations. Tregs expressing TIGIT selectively inhibit proinflammatory Th1 and Th17 cell responses. $Foxp3^+$ Tregs cells regulate immune responses and maintain self-tolerance. Recent work shows that Treg cells are comprised of many subpopulations with specialized regulatory functions. $Foxp3^+$ T cells expressing the coinhibitory molecule TIGIT are a distinct Treg cell subset that specifically suppresses proinflammatory T helper 1 (Th1) and Th17 cell, but not Th2 cell responses. TIGIT expression therefore identifies a Treg cell subset that demonstrates selectivity for suppression of Th1 and Th17 cell but not Th2 cell responses (Joller et al., 2014).

Controversially, abrogation of *Icosl* and with this a complete functional ICOS pathway led to both increase of chemokine receptor CXCR3 expressing Tregs and ICOS expressing Tregs. Kornete et al. observed that $ICOS^{+}$ Tregs preferentially expressed CXCR3 in the pancreatic lymph node of prediabetic NOD mice, and this expression gave them a better migratory ability to home to β-islets (Kornete et al., 2015). This discovery highlighted the crucial role of ICOS, as *Cxcr3*$^{-/-}$ NOD mice developed diabetes earlier than WT mice due to a decreased potential of these *Cxcr3*$^{-/-}$ Treg cells to migrate from pancreatic lymph nodes to β-islets (Yamada et al., 2012). Indeed, ICOS is indispensable for the expression of CXCR3 on Tregs. Tregs from *Icos*$^{-/-}$ NOD mice only express limited level of CXCR3 (Kornete et al., 2015). These CXCR3-expressing $ICOS^{+}$ Tregs showed a Th1-like phenotype, with an increased expression of T-bet and IFNγ as well as its receptor IFN-γR, which enables $ICOS^{+}$ Tregs to respond to IFNγ produced by effector T cells and provides Tregs with an enhanced inhibitory ability (Kornete et al., 2015). The *Tigit* mutation however does not impact on Treg frequency in the peripheral immune system.

*Th1/Th17 regulation in peripheral immune system of the NOD mouse*

IL-17 and IFNγ signalling synergistically contribute to the development of diabetes in NOD mice (Kuriya et al., 2013). Abrogation of the ICOS pathway decreases the frequency of CD226 expressing $CD4^{+}$ T cells and introduces a bias in IL-17/ IFNγ expressing cells that may well contribute to the increase in autoimmunity (Shapiro et al., 2020). Similar finding was made for the *Tigit* mutation. Interestingly, in the double *Icosl*/*Tigit* knockout, the bias in IL-17/ IFNγ expressing cells is reverted. The mechanisms to these observations remain to be elucidated.

*Limits of the present study and perspectives*

CD226 promotes, while TIGIT inhibits, $CD4^{+}$ T-cell proliferation and differentiation into a Th1 phenotype, as well as $CD8^{+}$ T-cell and NK cell cytotoxicity. Our study opens the perspective to functional analysis in the absence of *Tigit* in the nonobese diabetic mouse. TIGIT is also expressed on NK cells, activated T-cells, memory T-cells. TIGIT shares its two ligands, CD155 (PVR) and CD112 (PVRL2, nectin-2), with CD226. TIGIT can inhibit immune responses by delivering inhibitory signals to effector T and NK cells, by inducing tolerogenic dendritic cells, and by enhancing the suppressive capacity of Tregs, exerts its immunosuppressive effects by competing with other counterparts, CD266 (DNAM-1) or CD96. The TIGIT/CD226 pathway has been linked to multiple autoimmune diseases. Our study/ model therefore open up multiple choices for studying the TIGIT pathway.

**Author contributions**

C.Bé. designed, performed and analysed most of the experiments, S.L. designed and performed experiments, U.C.R. analysed data and wrote the manuscript, C.Bo. supervised the work and recruited funding. C.Bé, S.L., C.Bo. edited the manuscript.

**Acknowledgments**

We are grateful to our colleagues at the Animal house facilities, Patrice Vallin and the cytometry facility for their valuable technical contributions, Carine Giovannageli (TACGENE, Paris, France) for the CrispR design. Funding for this project was obtained from INSERM, CNRS, Université Paris Cité, and the Association française contre les myopathies (AFM).

**Figure legends**

**Figure 1. Cytometric T-cells analysis in spleen of 24-30 week-old WT mice and ICOSL$^{-/-}$ NOD mice.**

(a-c) Frequency analysis: (a) Spleen cells from wild type (n=22) and ICOSL$^{-/-}$ (n=23) female NOD mice were evaluated by flow cytometry for expression of TCR, B220, CD4, CD8, FoxP3, CD226. (b) CD4$^{+}$Foxp3$^{+}$ T cells were evaluated for ICOS, TIGIT, PD1, CXCR3 expression. (c) CD4$^{+}$Foxp3$^{-}$CD226$^{+}$ T-cells were evaluated for Tbet, RORγT, ICOS and CXCR3 expression; and (WT n=17, ICOSL$^{-/-}$ n=19) IFNγ, IL-17 and TNF production upon 4 h PMA stimulation. (d) Clustering analysis on 4500 TCR$^{+}$ cells per mouse using female WT (n=4) and *Icosl*$^{-/-}$ (n=4) NOD mice with different algorithms (see gating strategy in supplementary data); heat maps show expression levels of TCR, CD4, CD8 and FOXP3. (e) Clustering analysis on 8000 CD4$^{+}$FOXP3$^{+}$ cells per WT (n=4) and *Icosl*$^{-/-}$ (n=4) NOD mice; results were concatenated and plotted using the t-distributed Stochastic Neighbor Embedding (tSNE) algorithm; heat maps show expression levels of ICOS, TIGIT, PD1 and CXCR3. (f) As in Figure 1e, 50,000 FOXP3$^{-}$ CD4$^{+}$ cells per female wild type (n=4) and *Icosl*$^{-/-}$ (n=4) NOD mice were used; heat maps show expression of CD226, ICOS and CXCR3; (g) frequencies of Tbet-, RORgt- IFNγ- and IL17-expressing FOXP3$^{-}$CD4$^{+}$CD226$^{+}$ T-cells are shown. (h) Graphical representation of TIGIT-expressing spleen cells among CD4$^{+}$FOXP3$^{+}$ and CD4$^{+}$ FOXP3$^{-}$ T-cells in wild type (n=10) and *Icosl*$^{-/-}$ (n=10) 24-30 week-old female NOD mice; SD, standard deviation.

**Figure 2: Analysis of pancreatic, nerve and muscle infiltrates.**

Pancreatic infiltrates were obtained from female WT NOD mice, nerve and muscle infiltrates from female *Icosl*$^{-/-}$ mice at 24-33 weeks of age. (a-d) Frequency analysis: Data are representative of a single experiment using pooled cells from pancreas (WT n=5, 24 weeks), cells from nerves (*Icosl* $^{-/-}$ n= 6, 28-30 weeks), cells from muscles (*Icosl* $^{-/-}$ n= 5, 28 weeks). (a) Infiltrates were evaluated for TCR, B220, CD4, CD8, FoxP3 and ICOS expression, (b) for ICOS, TIGIT and PD1 expression within the CD4$^{+}$FoxP3$^{+}$ T-, (c) the CD4$^{+}$FoxP3$^{-}$ T-, and (d) the CD8$^{+}$ T-cell subset. (e) Average frequency of the ICOS, TIGIT and PD1 expressing cells among CD4$^{+}$ FOXP3$^{+}$, CD4$^{+}$FOXP3$^{-}$ and CD8$^{+}$ T cells in infiltrates; pancreas n=20 females (4 experiments at 24 weeks using 5 mice each), nerves n= 20 females (3 experiments at 28-33 weeks using 5,6 and 9 mice), muscle n=8 females (2 experiments at 25-28 weeks using 5 and 3 mice); for details see supplementary data. (f) Graphical representation of the percentage of TIGIT-expressing cells among $CD4^+FOXP3^+$ and $CD4^+$ $FOXP3^-$ cells in tissue infiltrates pooling the 4, 3 and 2 experiments for pancreatic, nerve and muscle infiltrates, respectively. SD, standard deviation.

**Figure 3: Construction of the NOD *Tigit*$^{-/-}$ mouse strain and changes in the peripheral T cell composition in NOD TIGIT$^{-/-}$ mice.**

(a) Right: PCR product profile after CRISPR/Cas9 transgenesis on an 1% agarose gel. Sizes of PCR products were determined using the GeneRuler 1kb DNA ladder (Thermo Scientific), 1651 bp corresponds to the WT allele and 597 bp corresponds to the *Tigit* knockout allele. Left: Evalution of the expression of TIGIT by cytometry on splenic $TCR^+CD4^+Foxp3^+$ cells from NOD WT mice (30.3 %) compared to TIGIT$^{-/-}$ NOD mice (0.45 %).

(b) Cumulative incidence of spontaneous diabetes in *Tigit*$^{-/-}$ *versus* wild type mice determined using 28 *Tigit*$^{-/-}$ females, 24 *Tigit*$^{-/-}$ males, 28 WT females, 19 WT males (n=19).

(c) Spleen cells were evaluated by flow cytometry for expression of TCR, B220, CD4, CD8, FoxP3, and CD266;

(d) among $CD4^+Foxp3^+$ cells for expression of ICOS and CXCR3; and (e) among $CD4^+Foxp3^-$ $CD226^+$ T-cells for Tbet, RORγT, ICOS and CXCR3 expression, and IFNγ, IL-17 and TNF production upon four h of PMA stimulation. Data are representative of two independent experiments using wild type (n=12) and TIGIT$^{-/-}$ (n=13) female NOD mice at 12 weeks of age.

(f) Clustering analysis carried out on five mice of each strain using different algorithms (see gating strategy in supplementary data). Results (1933 $CD226^+CD4^+$ T cells per mouse) were concatenated and plotted using the t-distributed Stochastic Neighbor Embedding (tSNE) algorithm to visualise and compare Tbet and RORgt expression. The frequency for each population is indicated on the graph.

**Figure 4: Changes in the peripheral T cell composition of ICOSL$^{-/-}$ TIGIT$^{-/-}$ compared to ICOSL$^{-/-}$ mice**

a) PCR validation of the *Icosl* deletion in *Icosl*$^{-/-}$*Tigit* $^{-/-}$mice (WT 225 bp, *Icosl*$^{-/-}$ 300 bp).

b) PCR validation of the *Tigit* deletion in *Icosl*$^{-/-}$*Tigit* $^{-/-/-}$ mice.

c) Incidence of neuropathy of WT females (n=9), ICOSL$^{-/-}$ females (n= 12) ICOSL$^{-/-}$ TIGIT$^{-/-}$ females (n=23).

d) Frequency of different cell types among peripheral lymphocytes. The cells were gated on their expression of the TCR. CD4$^{+}$ and CD8$^{+}$ T cell populations were identified. Among the CD4$^{+}$ T cells, regulatory FOXP3$^{+}$ cells were determined, and among the CD4$^{+}$ FOXP3$^{-}$ cells CD226$^{+}$ effector cells were discerned. e) Frequency of regulatory T cells expressing the markers ICOS, TIGIT, PD1, CXCR3. f) Characterization of the transcription factors Tbet and RORgt in the CD4$^{+}$FOXP3$^{-}$CD226$^{+}$ T cell population and analysis of IL-17, IFNγ and TNF cytokine production upon PMA stimulation for 4 hours. Data are representative of three independent experiments using female ICOSL$^{-/-}$ (n= 13, n=9 for cytokines) and ICOSL$^{-/-}$ TIGIT$^{-/-}$ mice (n=17, n=11 for cytokines) at 24-30 weeks of age.

g) The clustering analysis was carried out on five mice of each strain using different algorithms (see gating strategy in supplementary data). T cell types (4547 CD226$^{+}$ CD4$^{+}$ T cells per mouse) were plotted using the t-distributed Stochastic Neighbor Embedding (tSNE) algorithm. Heat maps show expression levels of designated markers.

**Supplementary data Figure 1. Analysis of spleen and pancreatic T-cells in WT NOD mice at 12 weeks**.

(a-d) Flow cytometry of NOD wild type spleens (n=11). (a) Cells were analyzed for expression of TCR, B220, CD4, CD8, FoxP3, CD226; (b) ICOS, TIGIT, PD1 and CXCR3 were analysed on CD4$^{+}$Foxp3$^{+}$, (c) CD4$^{+}$Foxp3$^{-}$CD226$^{-}$, CD4$^{+}$Foxp3$^{-}$CD226$^{+}$ T-cells; (d) CD226, ICOS, CXCR3 and PD1 were analysed on CD8$^{+}$ T cells; (e) Gating strategy: selection of TCR$^{+}$B220$^{-}$ cells, then of CD8$^{+}$ T-cells and CD4$^{+}$ T-cells and, among CD4$^{+}$ T-cells, of CD4$^{+}$FoxP3$^{+}$ T-cells and CD4$^{+}$FoxP3$^{-}$ T-cells, among CD4$^{+}$FoxP3$^{-}$ T-cells, of CD4$^{+}$FoxP3$^{-}$ CD226$^{+}$ and CD226$^{-}$ T-cells. (f) Expression of Tbet and RORgt and analysis of IL-17, IFNγ and TNF cytokine production upon 4 h PMA stimulation of CD4$^{+}$FOXP3$^{-}$CD226$^{-}$ *versus* CD4$^{+}$FOXP3$^{-}$ CD226$^{+}$ T cells and (g) CD8$^{+}$ T-cells. (h) Graph showing the cellular distribution and frequency of ICOS, TIGIT and PD1 expressing cells among CD4$^{+}$ FOXP3$^{+}$, CD4$^{+}$FOXP3$^{-}$ and CD8$^{+}$ T cells in pancreatic infiltrates of female NOD mice at 12 weeks; means of four independent experiments that pooled five pancreata each; SD, standard deviation.

**Supplementary data Figure 2: Frequencies of the peripheral T cells in wild type and ICOSL$^{-/-}$ 16-week-old NOD mice.**

(a) Spleen cells from wild type (n=6) and ICOSL$^{-/-}$ (n=5) female NOD mice were evaluated by flow cytometry for expression of TCR, B220, CD4, CD8, FoxP3, CD226; (b) CD4$^{+}$Foxp3$^{+}$ T cells were evaluated for ICOS, TIGIT, PD1, CXCR3 expression; (c) CD4$^{+}$Foxp3$^{-}$CD226$^{+}$ and CD4$^{+}$Foxp3$^{-}$CD226$^{-}$ T-cells were evaluated for ICOS and CXCR3 expression; (d) CD8$^{+}$ T-cells were evaluated for CD226, ICOS, CXCR3 and PD1 expression; (e) CD4$^{+}$Foxp3$^{-}$CD226$^{-}$ and CD4$^{+}$Foxp3$^{-}$CD226$^{+}$ T-cells were evaluated for Tbet, RORγT, ICOS and CXCR3 expression, and IFNγ, IL-17 and TNF production upon 4 h PMA stimulation; (f) CD8$^{+}$ T-cells were evaluated for CD226, Tbet, RORγt, ICOS, CXCR3 expression and and IFNγ, IL-17 and TNFproduction upon 4 h PMA stimulation. Data are representative of a single experiment using female NOD WT (n=6) and NOD ICOSL$^{-/-}$ mice at 16 weeks (n=5) for phenotyping and two experiments for transcription factors: NOD WT n=11 (n=5 for cytokines) and for NOD ICOSL$^{-/-}$n= 6 (n= 3 for cytokines).

**Supplementary data Figure 3: Analysis of the peripheral T cell composition of mice in NOD TIGIT$^{-/-}$ mice weeks at 16 weeks of age**

a) Frequency of different cell types among peripheral lymphocytes. The cells were gated on their expression of the TCR. T CD4$^{+}$ and T CD8$^{+}$ populations were identified. Then among the population CD4$^{+}$, the regulating population T expressing FOXP3 was determined and among the effector population CD4$^{+}$ FOXP3$^{-}$, the activating population CD226$^{+}$ was identified.

b) Among the CD4$^{+}$FoxP3$^{+}$ cells, the markers ICOS, CXCR3 were examined.

c) Expression of transcription factors Tbet and RORgt in the CD4$^{+}$FOXP3$^{-}$CD226$^{-}$ and study of their ability to produce IL-17, IFNγ and TNF after PMA stimulation for 4 hours.

d) Expression of transcription factors Tbet and RORgt in the CD4$^{+}$FOXP3$^{-}$CD226$^{+}$ cells and study of their ability to produce IL-17, IFNγ and TNF after PMA stimulation for 4 hours.

e) Expression of transcription factors Tbet and RORgt in the CD8$^{+}$ cells and study of their ability to produce IL-17, IFNγ and TNF after PMA stimulation for 4 hours.

Data are representative of a single experiment using female NOD WT (n=5) and NOD TIGIT$^{-/-}$ 16 weeks (n=4).

**Supplementary data Figure 4: flow cytometry gating and analysis strategy**

a) Data acquisition was performed using a BD Biosciences LSR-Fortessa. Results were analysed with FlowJo analysis software V10.6.1. Data were cleaned using the FlowJo pluging FlowAI, developed by Gianni Monaco et al., to check parameters over time and to look for deviations outside the statistical norm. Gating was performed as shown. Populations of interest $CD4^+$, $CD4^+FOXP3^+$ or $FOXP3^-CD226^+$ T cells were concatenated at equal number of cells and equal number of mice. tSNE , HeatMaps and flowSOM plugings were used to visualize the clustering.

FIG1
V17-1

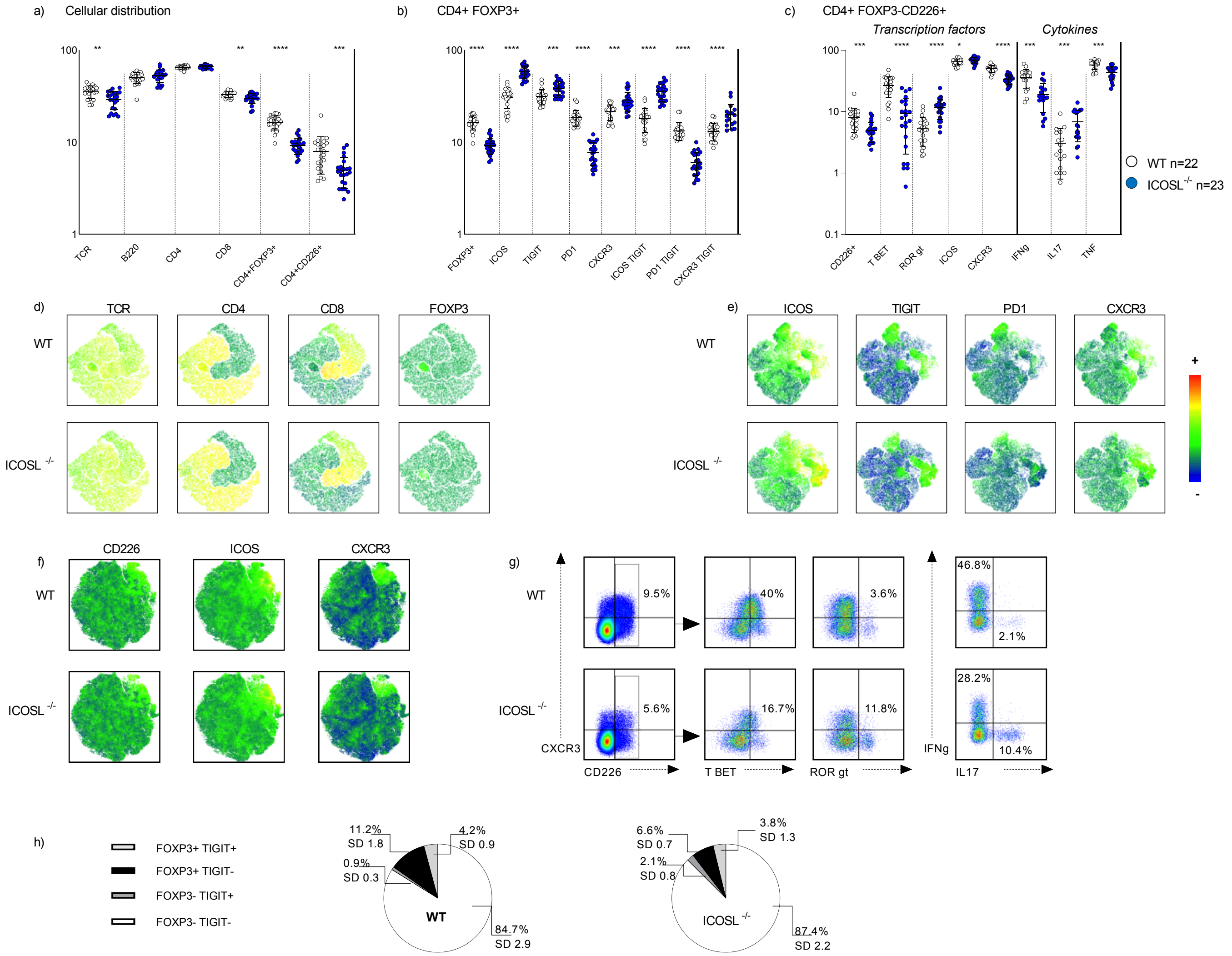

FIG2
V17-1

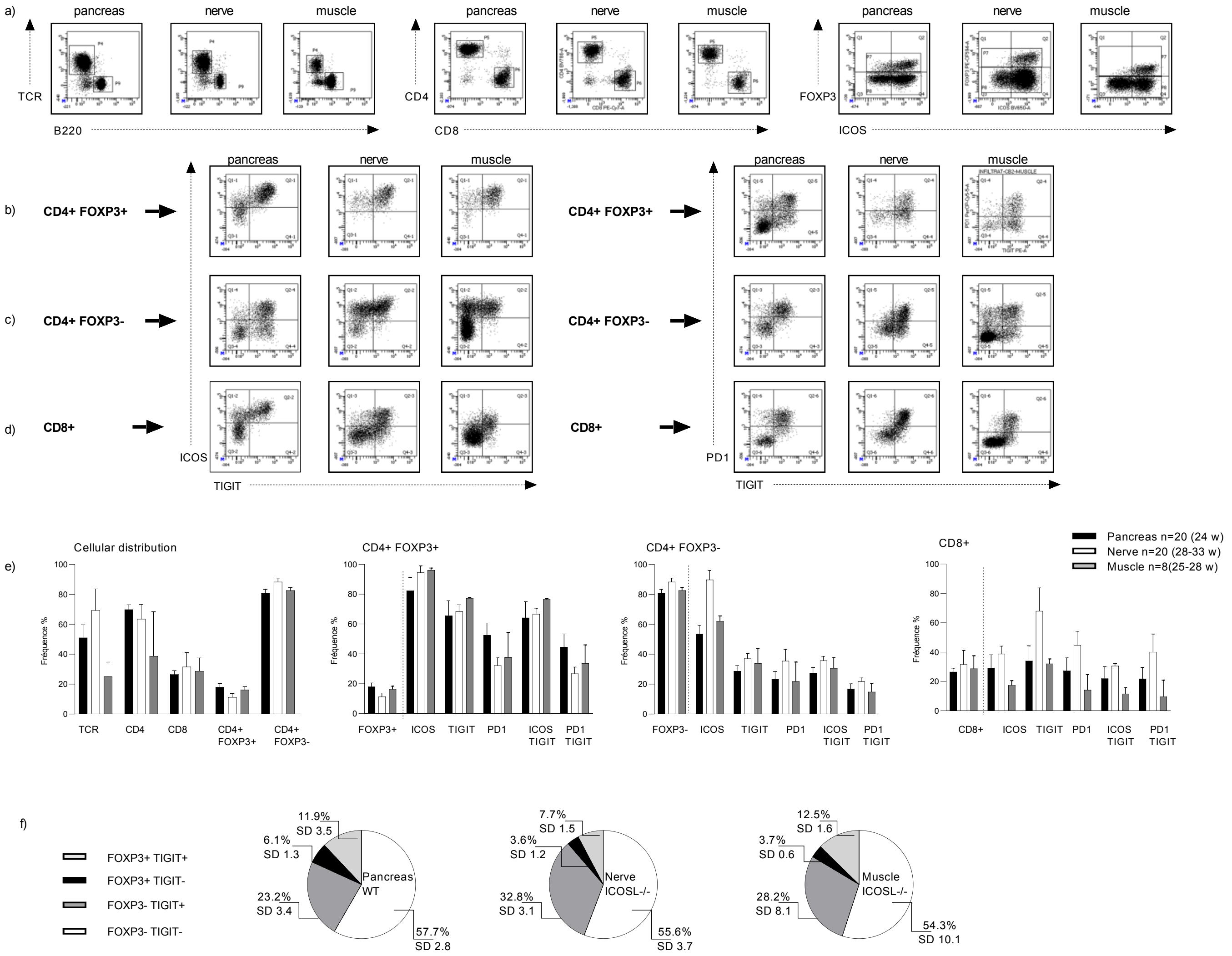

**FIG3**
**V17-1**

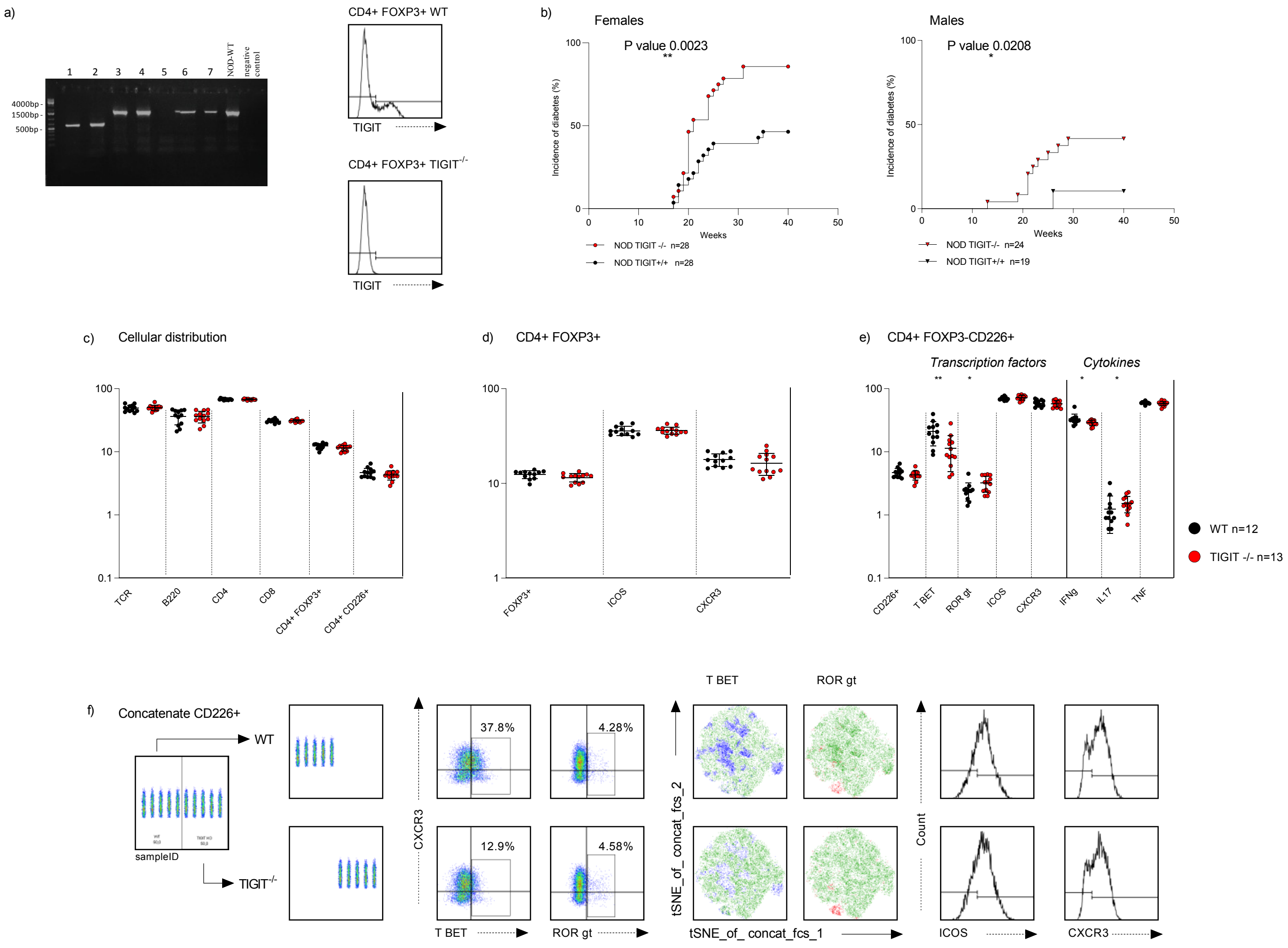

FIG4
V17-1

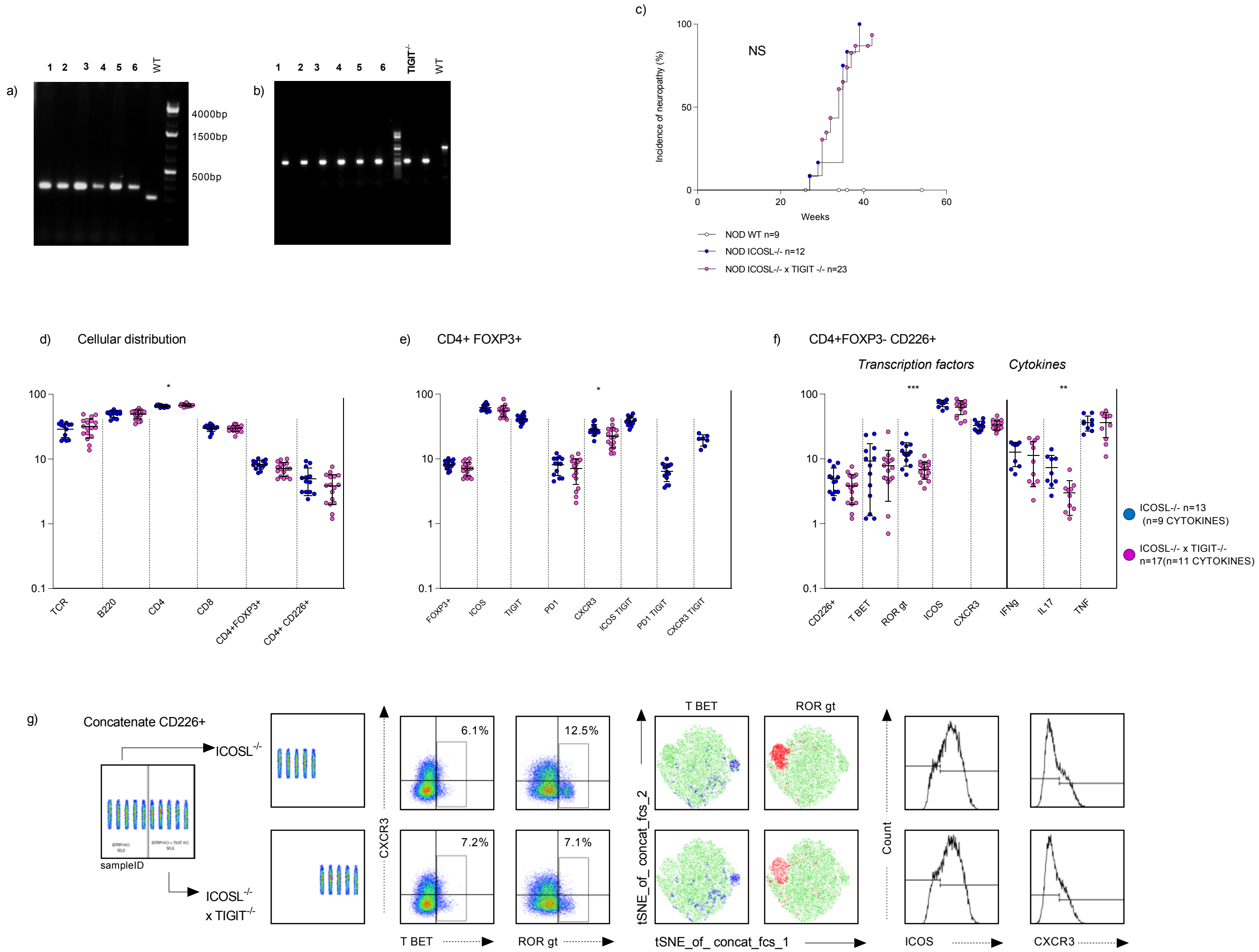

a)
1 2 3 4 5 6 WT
4000bp
1500bp
500bp
b)
1 2 3 4 5 6 TIGIT-/- WT
c)
NS
Incidence of neuropathy (%)
Weeks
NOD WT n=9
NOD ICOSL-/- n=12
NOD ICOSL-/- x TIGIT -/- n=23
d) Cellular distribution
TCR
B220
CD4
CD8
CD4+FOXP3+
CD4+ CD226+
e) CD4+ FOXP3+
FOXP3+
ICOS
TIGIT
PD1
CXCR3
ICOS TIGIT
PD1 TIGIT
CXCR3 TIGIT
f) CD4+FOXP3- CD226+
Transcription factors
Cytokines
CD226+
T BET
ROR gt
ICOS
CXCR3
IFNg
IL17
TNF
ICOSL-/- n=13
(n=9 CYTOKINES)
ICOSL-/- x TIGIT-/-
n=17(n=11 CYTOKINES)
g) Concatenate CD226+
ICOSL-/-
sampleID
ICOSL-/- x TIGIT-/-
CXCR3
6.1%
12.5%
7.2%
7.1%
T BET
ROR gt
tSNE_of_concat_fcs_2
tSNE_of_ concat_fcs_1
Count
ICOS
CXCR3

**Data sup FIG1**
**V17-1**

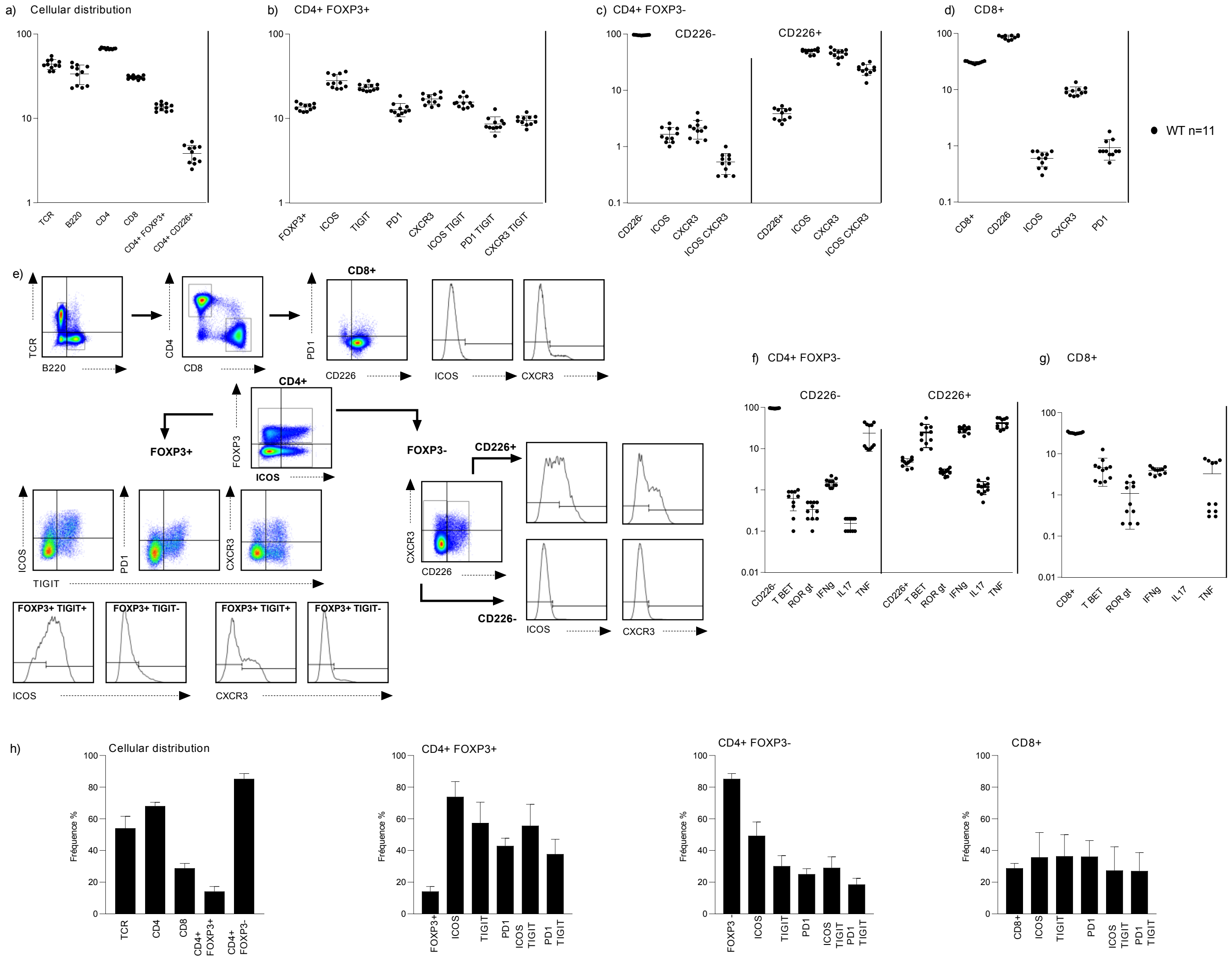

**Data sup FIG2**
**V17-1**

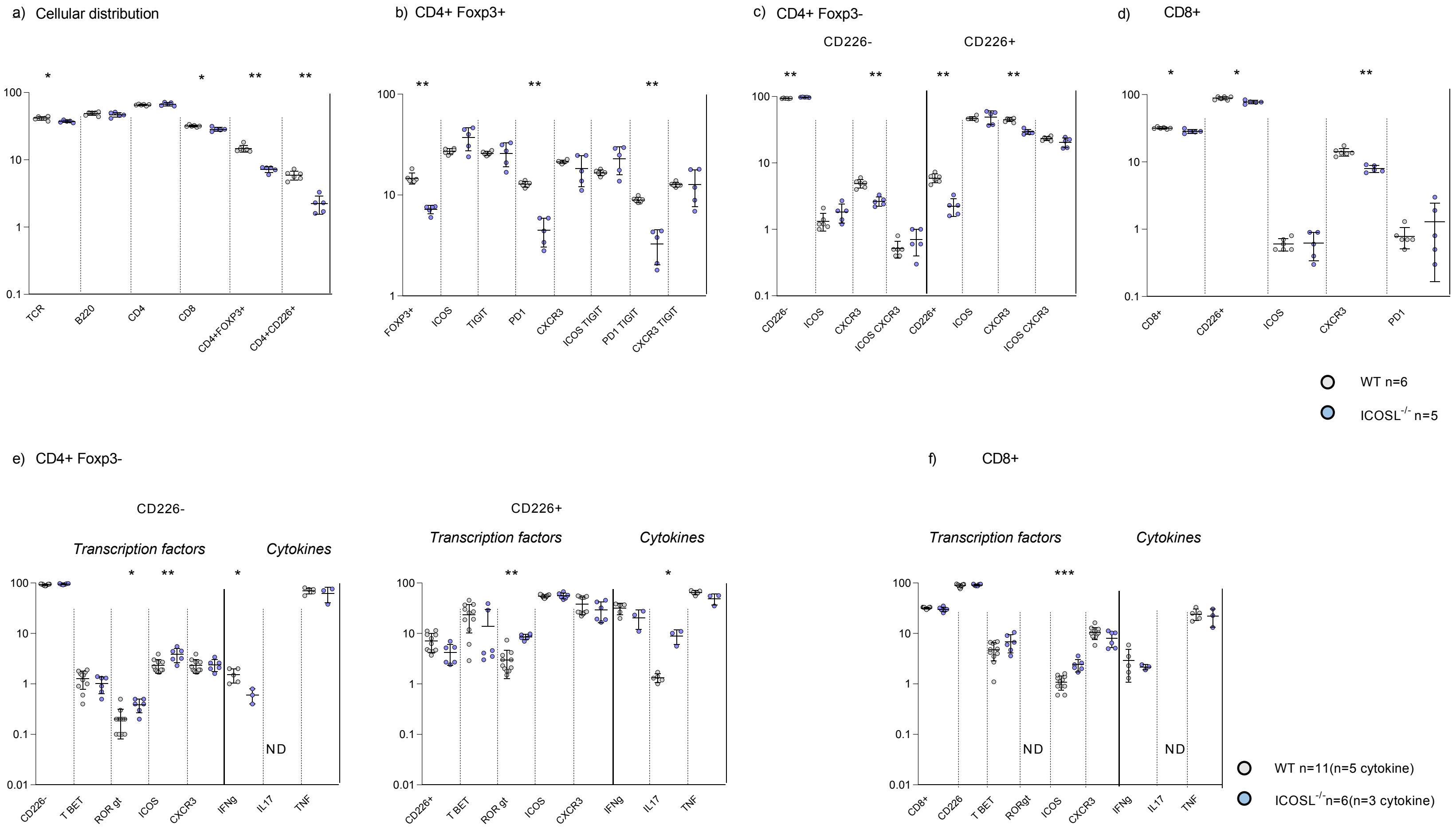

**Data sup FIG3**
**V17-1**

a) Cellular distribution

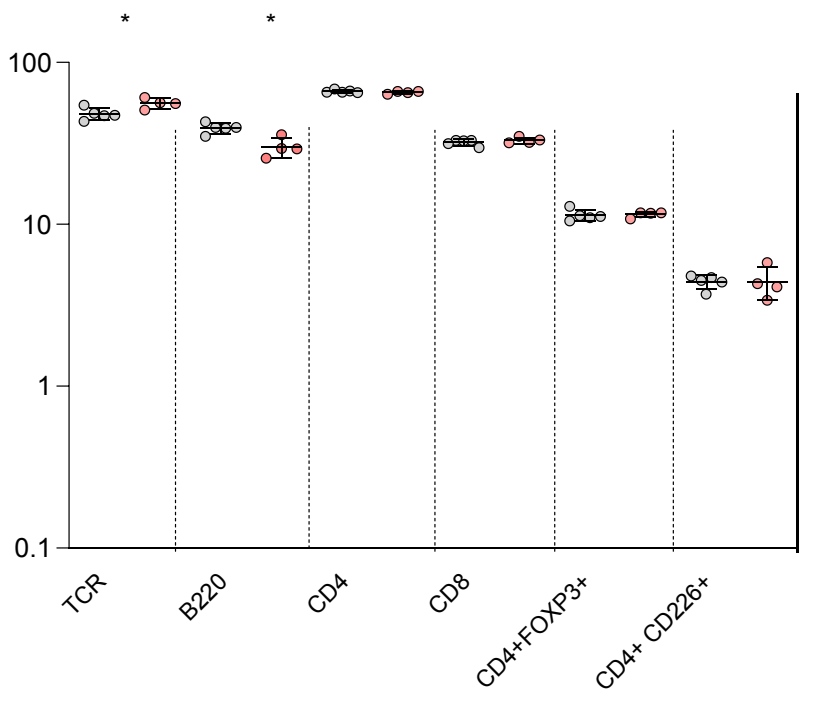

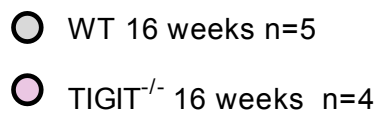

b) CD4+ Foxp3+

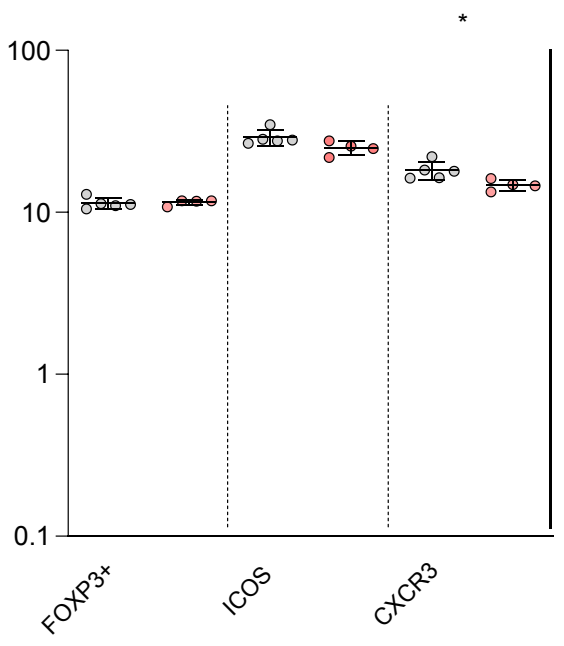

c) CD4+FOXP3-CD226-

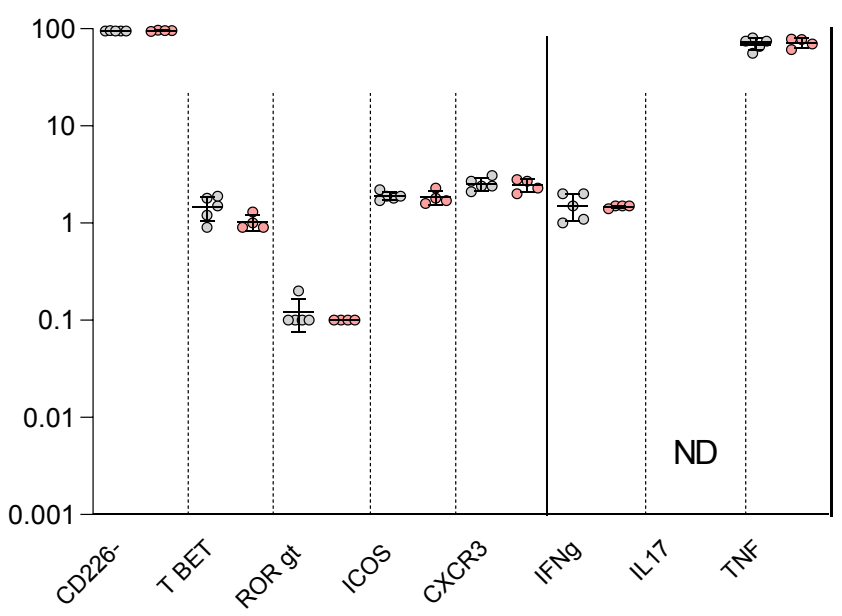

d) CD4+ FOXP3- CD226+

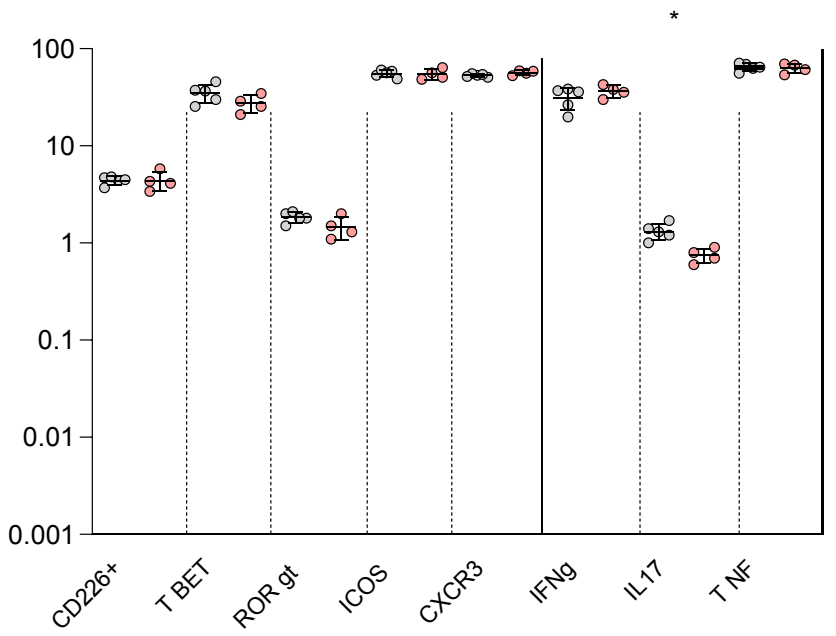

e) CD8+

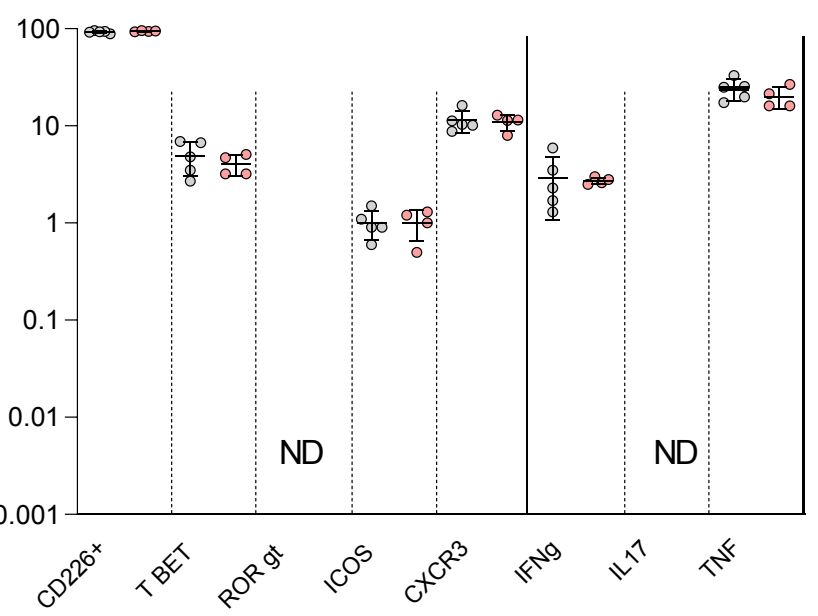

**Data sup FIG4**
**V17-1**

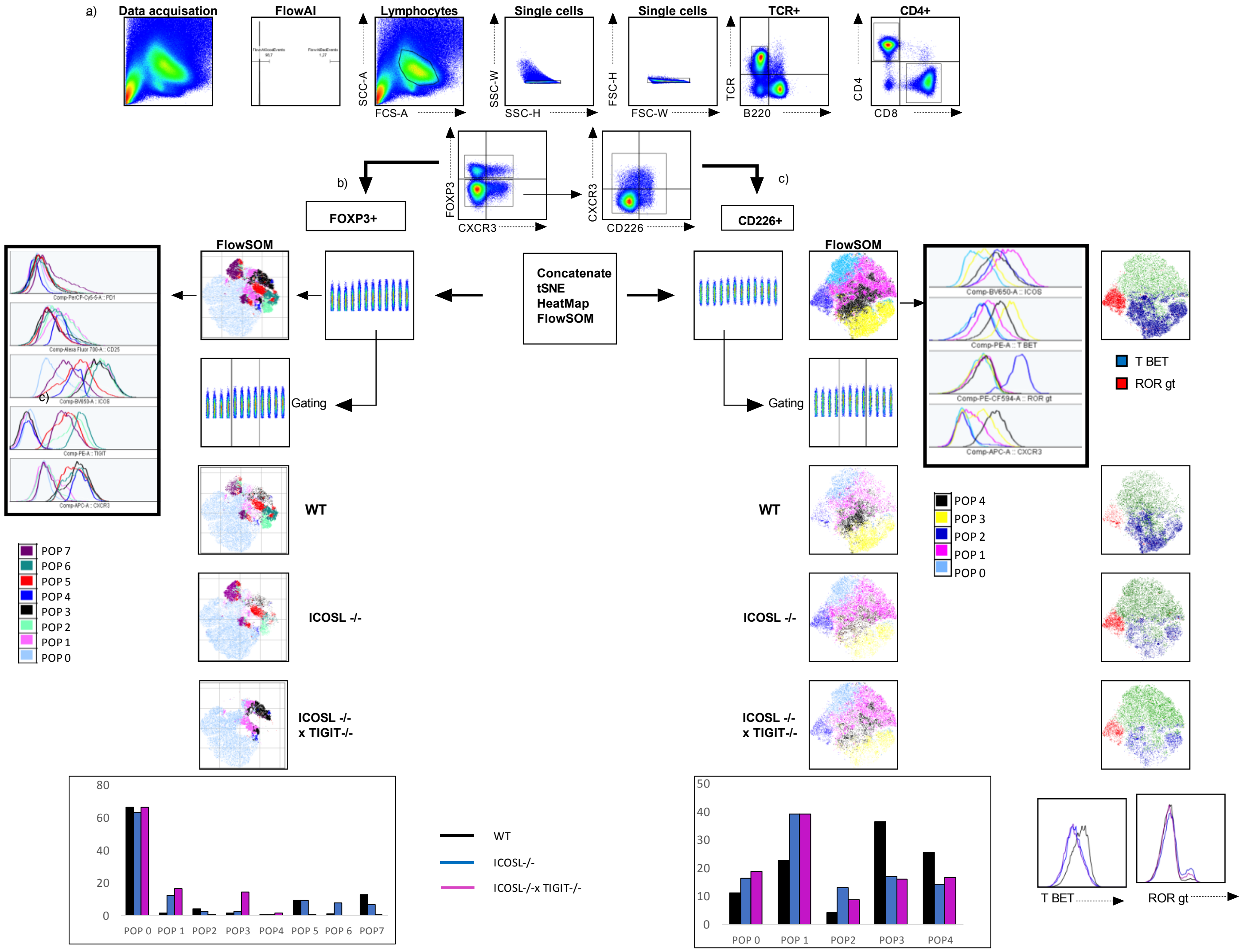